\documentclass[10pt,twocolumn,aps,prc,longbibliography,nofootinbib,floatfix]{revtex4-1}
\usepackage{graphicx,bm,xcolor, bbold}
\usepackage[normalem]{ulem}
\usepackage{hyperref}
\hypersetup{colorlinks, linkcolor={blue},citecolor={blue},urlcolor={blue}}
\usepackage{graphicx}
\usepackage{amsmath,amssymb,amsfonts}
\usepackage{bm}
\usepackage{color}
\graphicspath{fig}
\usepackage{epstopdf}
\usepackage{multirow}
\usepackage{scalerel}

\newcommand{\new}[1]{{\color{red}{#1}}}

\newcommand{\remove}[1]{\ifmmode\mbox{\new{\sout{$#1$}}}\else\new{\sout{#1}}\fi}

\newcommand{\vect}[1]{\bm{\mathrm{#1}}}

\newcommand{\Jv}{\vect{J}}
\newcommand{\Iv}{\vect{I}}
\newcommand{\av}{\vect{a}}
\newcommand{\bv}{\vect{b}}

\newcommand{\jv}{\vect{j}}
\newcommand{\kv}{\vect{k}}

\newcommand{\nablav}{\vect{\nabla}}
\newcommand{\pv}{\vect{p}}

\newcommand{\qv}{\vect{q}}
\newcommand{\Qv}{\vect{Q}}
\newcommand{\rv}{\vect{r}}
\newcommand{\rhov}{\vect{\rho}}
\newcommand{\vv}{\vect{v}}

\newcommand{\xv}{\vect{x}}
\newcommand{\xvhat}{\hat{\vect{x}}}
\newcommand{\yvhat}{\hat{\vect{y}}}
\newcommand{\zvhat}{\hat{\vect{z}}}

\newcommand{\calH}{\mathcal{H}}

\newcommand{\nv}{\vect{n}}
\newcommand{\fm}{\;\text{fm}}
\newcommand{\MeV}{\;\text{MeV}}

\begin{document}
\title{Superfluid fraction in the crystal phase of the inner crust of neutron stars}
\author{Giorgio Almirante}
\altaffiliation[Present address: ]{Center for Computational Sciences, University of Tsukuba, Tsukuba 305-8577, Japan}
\email[\\]{giorgio.almirante@nucl.ph.tsukuba.ac.jp}
\affiliation{Universit\'e Paris-Saclay, CNRS/IN2P3, IJCLab, 91405 Orsay, France}
\author{Theodora Kaskitsi}
\email{theodora.kaskitsi@universite-paris-saclay.fr}
\affiliation{Universit\'e Paris-Saclay, CNRS/IN2P3, IJCLab, 91405 Orsay, France}
\author{Michael Urban}
\email{michael.urban@ijclab.in2p3.fr}
\affiliation{Universit\'e Paris-Saclay, CNRS/IN2P3, IJCLab, 91405 Orsay, France}
\begin{abstract}
In the most extended layer of the inner crust of neutron stars, nuclear matter is believed to form a crystal of clusters immersed in a superfluid neutron gas.
Here we analyze this phase of matter within fully self-consistent Hartree-Fock-Bogoliubov calculations using Skyrme-type energy density functionals for the mean field and a separable interaction in the pairing channel.
The periodicity of the lattice is taken into account using Bloch boundary conditions, in order to describe the interplay between band structure and superfluidity. A relative flow between the clusters and the surrounding neutron gas is introduced in a time-independent way. As a consequence, the complex order parameter develops a phase, and in the rest frame of the superfluid one finds a counterflow between neutrons inside and outside the clusters. The neutron superfluid fraction is computed from the resulting current. Our results indicate that at densities above $0.03\fm^{-3}$, more than $90\%$ of the neutrons are effectively superfluid, independently of the detailed choice of the interaction, cluster charge, and lattice geometry. This fraction is only slightly lower than the one obtained recently within linear response theory on top of the Bardeen-Cooper-Schrieffer approximation, and it approaches the hydrodynamic limit for strong pairing. As a consequence, it is likely that the inner crust alone can provide a sufficient superfluid angular momentum reservoir to explain pulsar glitches.
\end{abstract}
\maketitle
%
\section{Introduction}
In the inner crust of neutron stars, protons and neutrons are believed to form clusters that are arranged in a periodic lattice, immersed in a superfluid neutron gas and a relativistic electron gas ensuring charge neutrality and $\beta$-equilibrium \cite{Chamel08}.
The presence of the superfluid neutron gas in the inner crust has important effects on the star's hydrodynamic and thermodynamic behavior \cite{Page12}, on its shear modes \cite{Andersson21}, on the equilibrium configuration of rotating stars \cite{Prix02}, and especially on pulsar glitches \cite{Anderson75}.
The neutron superfluid fraction is a central microscopic parameter for glitch models \cite{Carter06A,Antonelli22}.

At zero temperature, a uniform one-component superfluid (independently of whether it consists of bosons or paired fermions) has a superfluid fraction of 100\%~\cite{Leggett98}. 
This remains true even in weakly coupled fermionic systems, although pairing affects in this case only particles near the Fermi surface. The reason why the superfluid fraction can still be 100\% is that the superfluid density is the long-wavelength limit of a response function \cite{Schrieffer}, which is dominated by states near the Fermi surface. However, in non-uniform systems such as the inner crust, this fraction is necessarily reduced \cite{Leggett98}.

The reduction of the superfluid density $\rho_S$ can be related to the ``entrainment'', i.e., a non-dissipative coupling between superfluid neutrons and the protons \cite{Prix02}.
In a uniform neutron-proton mixture, entrainment arises from momentum dependence of the neutron-proton interaction (Andreev–Bashkin effect \cite{Andreev75}).
In the presence of a periodic lattice, there is an additional effect due to Bragg scattering \cite{Chamel12A}, analogous to band gaps in electron systems \cite{Ashcroft}.
Modeling this phenomenon requires sophisticated band-structure calculations for the neutrons \cite{Chamel05,Chamel06,Chamel12A,Kashiwaba19,Sekizawa22}.
Systematic calculations for the crystal phase \cite{Chamel12A} predicted a very strong entrainment and thus a small superfluid fraction.
This presents a tension with glitch observations \cite{Chamel13}, unless one abandons the
assumption that glitches originate solely in the crust \cite{Andersson12}.

This prediction of band theory is at variance with the result of an alternative approach based on superfluid hydrodynamics \cite{Martin16} (extending calculations for an isolated cluster \cite{Sedrakian96,Magierski04,Magierski04a,Magierski04b} to a periodic lattice), which predicted a much weaker entrainment and hence a much larger superfluid fraction.
If accurate, this would help explain the observed glitches in the Vela pulsar.

However, both approaches have their limitations. The hydrodynamic approach is valid in the limit of strong pairing, when Cooper pairs are smaller than clusters and lattice spacing \cite{Martin16}, but this condition is not fulfilled in the inner crust.
Conversely, normal band theory calculations \cite{Chamel05,Chamel06,Chamel12A} neglect the pairing gap $\Delta$.
First band-structure calculations including pairing within the Bardeen-Cooper-Schrieffer (BCS) approximation \cite{Carter05A} found that the gap has almost no effect on the superfluid fraction \cite{Chamel25A}.
This is in contradiction with Hartree-Fock-Bogoliubov (HFB) calculations in a sinusoidal potential in one direction \cite{Watanabe17} and with our recent HFB calculations in the rod phase \cite{Almirante24A} which found an increase of the superfluid fraction with increasing pairing gap.
In fact, from ultracold atoms, it is known that HFB recovers hydrodynamic behavior in the limit of strong pairing \cite{Grasso05,Tonini06}.
As a possible explanation for this discrepancy, a study within a toy model example questioned the validity of the BCS approximation \cite{Minami22}. However, currents computed in linear response approach ideal-fluid behavior in the limit of strong pairing even when built on top of the BCS approximation \cite{Migdal59,Thouless62,Urban03}.
Recently, we showed in \cite{Almirante25} that the issue in \cite{Carter05A,Chamel25A} was not the BCS approximation to the ground state, but the neglect of a term in the linear response, known as the ``geometric contribution'' in the fields of cold-atom systems \cite{Peotta15,Liang17} and multiband superconductors \cite{Iskin24,Jiang24,Iskin25}.
Once this contribution is included, the superfluid density increases strongly with increasing pairing gap.

However, the linear-response expression in \cite{Almirante25} assumes a constant gap, does not include the effective mass, and ignores the change of the gap, i.e., its phase, induced by the flow. 
This phase appears in HFB \cite{Thouless62,Almirante24,Almirante24A} and hydrodynamics \cite{Martin16} and generates an additional term in the linear-response expression \cite{Migdal59}, which is necessary to fulfil the continuity equation, but which was not included in \cite{Almirante25}.
The continuity equation is actually what leads to the Leggett upper bound of the superfluid fraction \cite{Leggett98} in the case of density modulations in only one direction. According to \cite{Saslow12}, the superfluid hydrodynamics result is the generalization of this upper bound to three dimensions (3D), and it is actually well-defined for any density profile \cite{Saslow76,Saslow12,Rabec25} and not only for the schematic ones as in \cite{Martin16}.

Full dynamical pairing therefore requires solving the HFB equations. For the description of a stationary flow, this can be done in a time independent way by using a Galilean transformation, as in our previous works on the pasta phases \cite{Almirante24,Almirante24A}. Time dependent HFB studies including band structure exist so far only for the slab phase \cite{Yoshimura24,Yoshimura25}.

In this work, we perform fully self-consistent HFB calculations for the crystal phase, extending our studies in the rod (``spaghetti'') \cite{Almirante24A} and slab (``lasagna'') \cite{Almirante24} phases, treating the neutrons as superfluid and the protons as normal fluid. 
Section \ref{sec:formalism} gives a brief summary of the formalism, in particular of the interactions and of the definition of the superfluid density in a two-fluid description of a relative flow between clusters and superfluid. Section~\ref{sec:results} shows HFB results with and without relative flow and compares them with the approximate linear response on top of BCS and with superfluid hydrodynamics. Conclusions are given in Sec.~\ref{sec:conclusion} and technical details in the Appendices.

\section{Formalism} \label{sec:formalism}
In this section, we briefly recall the general formalism, which is practically the same as in our previous works about the pasta phases \cite{Almirante24,Almirante24A}, and discuss some extensions. The calculations will then be performed for the crystal phase, in the simple cubic and body-centered cubic lattices. Details about periodicity and reciprocal lattices are given in Appendices \ref{appA} and \ref{appB}.

\subsection{Hamiltonian}
To calculate the microscopic properties of the inner crust of neutron stars, we consider a system composed of superfluid neutrons and normal protons arranged in a periodic lattice. Accordingly, we perform Hartree-Fock-Bogoliubov (HFB) calculations for neutrons and Hartree-Fock (HF) calculations for protons. Electrons are also present in the system, they are treated as a uniform, constant background that ensures overall charge neutrality.

For the mean-field description, we employ Skyrme-type energy-density functionals. Specifically, we use the parameterizations SLy4 \cite{Chabanat97,Chabanat98} from the Saclay-Lyon family as well as BSk24 \cite{Chamel09,Chamel13A} from the Bruxelles-Montreal family.

For simplicity, we omit the spin-orbit term in our calculations.
In the case of protons, we account for the Coulomb interaction, including the exchange term using the Slater approximation (this is applied only with SLy4, as BSk24 does not include this term in its parameter fitting).

By taking the functional derivatives of the energy density with respect to number density $\rho_q$, kinetic energy density $\tau_q$ and momentum density $\jv_q$ (where $q = n,p$), one gets the effective masses $m^*_q$, mean-field potentials $U_q$ and the momentum dependent terms $\Iv_q$ required by Galilean invariance.\footnote{See \cite{Almirante24} for details, except that there the symbol $\Jv$ was used instead of the symbol $\Iv$ which we are using here following \cite{Chamel19,Allard21}.}
Then the mean-field Hamiltonian reads in momentum space for each species (we drop the index $q$ for better readability)
\begin{equation}
\label{eq:hmeanfield}
     h_{\kv\kv'} =
     \kv\cdot\kv'\Big(\frac{\hbar^2}{2m^*}\Big)_{\kv-\kv'} \hspace{-1mm}+
     U_{\kv-\kv'}
     - (\kv+\kv') \cdot \Iv_{\kv-\kv'}\,.
\end{equation}
A flow can be induced in the system by replacing in the HFB equations the mean-field Hamiltonian \eqref{eq:hmeanfield} by 
\begin{equation}
\label{eq:Hpv}
    h_{\kv\kv'}(\vv)=h_{\kv\kv'}
    - \hbar\kv\cdot\vv \delta_{\kv\kv'}\,.
\end{equation}
As a consequence of the last term, non-vanishing momentum densities $\jv_q$ will arise (see \cite{Almirante24} for details on this Galilean boost). It has to be noticed that since protons are normal fluid, Galilean invariance ensures that $\vv$ is the proton velocity.
\vspace{3mm}\\
For the pairing field, a non-local interaction written in separable form is implemented, i.e.
\begin{equation}
    V_{\kv_1\kv_2\kv_3\kv_4}^{\text{pair}} = -g
    \,
    f\Big(\frac{|\kv_1+\kv_2|}{2}\Big)
    f\Big(\frac{|\kv_3+\kv_4|}{2}\Big)
    \delta_{\kv_1-\kv_2,\kv_3-\kv_4}\,,
\end{equation}
The form factors $f(k)$ are taken to be Gaussians 
\begin{equation}
    f(k)=e^{-k^2/k_0^2}\,.
\end{equation}
The coupling constant $g$ and the Gaussian width $k_0$ had been fitted on the $V_{\text{low-}k}$ interaction in \cite{Martin14}. Given the above interaction, the pairing gap also results non-local, which in momentum space reads
\begin{equation} \label{eq:gapk}
    \Delta_{\kv\kv'} = g \hspace{1mm}
    f\Big(\frac{|\kv+\kv'|}{2}\Big) \sum_{\pv\pv'}
    f\Big(\frac{|\pv+\pv'|}{2}\Big) \kappa_{\pv\pv'}
    \delta_{\kv-\kv',\pv-\pv'}\,,
\end{equation}
where $\kappa_{\pv\pv'}$ is the anomalous density matrix. The non-local pairing gap can be rewritten in Wigner (phase-space) representation, namely
\begin{equation} \label{eq:gapsep}
    \Delta(\Qv,\xv)= f(\Qv) \Delta_0(\xv)\,,
\end{equation}
with $\xv$ the Cooper pair center-of-mass (c.o.m.) position and $\Qv$ its relative momentum. The fact that the form factor $f(k)$ is real implies that the above expression can be rearranged as
\begin{equation}\label{eq:DeltaQx}
    \Delta(\Qv,\xv)= f(\Qv) |\Delta_0(\xv)| e^{i\phi(\xv)}\,,
\end{equation}
where $\phi$ is the phase of the pairing field. Notice that the phase $\phi$ is a function of the pair c.o.m. position only, because of the separable form of the pairing interaction.

The matrices $h(\vv)$ and $\Delta$ are then combined in a block matrix
\begin{equation}\label{eq:HFBH}
 \calH =
 \begin{pmatrix}
     h(\vv)-\mu & -\Delta \\
     -\Delta^\dagger & -\bar{h}(\vv)+\mu
 \end{pmatrix} ,
\end{equation}
where $\mu$ is the chemical potential and $\bar{h}_{\kv \kv'}=h_{-\kv' -\kv}$. Then one has to solve the eigenvalue problem
\begin{equation}\label{eq:HFBeigen}
    \calH\begin{pmatrix}U^*_\alpha\\-V^*_\alpha\end{pmatrix} 
    = E_\alpha \begin{pmatrix}U^*_\alpha\\-V^*_\alpha\end{pmatrix} .
\end{equation}
This construction is implemented in momentum space, imposing Bloch boundary conditions. This means that the momentum $\kv$ (and analogously $\kv'$) is decomposed into an integer label (which indicates the reciprocal lattice points), and a continuous Bloch momentum $\kv_b$, that is restricted to the first Brillouin zone (BZ). In this representation, the matrices $h$ and $\Delta$ are diagonal in $\kv_b$ and the diagonalization has to be performed on the reciprocal lattice. In Appendices \ref{appA} and \ref{appB} we discuss this in detail. Once the band structure and its corresponding eigenvectors are computed, one can get all the relevant densities and currents as explained is Appendix \ref{appA} and iterate this procedure until self-consistency is reached.

\subsection{Two-fluid flow} \label{subsec:two-fluid-flow}
In systems in which superfluid and normal neutrons and protons coexist, the flow can be described by the formalism due to Andreev and Bashkin \cite{Andreev75}. Since we assume that protons are not superfluid, the Andreev-Bashkin relations for the particle currents $\rhov_q$ simplify to those of a two-fluid model,
\begin{align} 
    \rhov_n &=(\rho_n-\rho_S)\vv_N +
    \rho_S \vv_S\,, \label{AB4}\\
    \rhov_p &=\rho_p \vv_N\,, \label{AB4protons}
\end{align}
with $\rho_n$ and $\rho_p$ the number densities of neutrons and protons, $\vv_N$ the velocity of the normal fluid, and $\rho_S$ and $\vv_S$ the neutron superfluid density and neutron superfluid velocity, respectively. In our case, the normal-fluid velocity $\vv_N$ coincides with the velocity of the clusters and with the proton velocity $\vv$ in the mean-field Hamiltonian \eqref{eq:Hpv}.

Since our system is inhomogeneous, the relations \eqref{AB4} and \eqref{AB4protons} have to be understood on a ``coarse-grained'' scale. Thus, $\rho_n$ and $\rho_p$ should be replaced with their cell averages $\langle\rho_n\rangle$ and $\langle\rho_p\rangle$. The neutron superfluid velocity $\vv_S$ is given by the gradient of the coarse-grained phase $\langle\phi\rangle$ of the pairing field \cite{Pethick10}, namely
\begin{equation} \label{eq:supvel1}
    \vv_S = \int_V \frac{d^3x}{V}
    \frac{\hbar}{2m}\nablav\phi\,,
\end{equation}
where $V$ is the volume of the lattice primitive cell and $\phi$ is the microscopic (not averaged) phase. Notice that $\rho_S$ is a tensor, i.e., it depends on the direction of the flow. If the system has cubic symmetry, one has $\rho_S^{ij}=\rho_S \delta^{ij}$ \cite{Carter06} and the scalar superfluid density is $\rho_S=\sum_i\rho_S^{ii}/3$.

Working in the frame in which the phase of the pairing field is periodic, Eq.~\eqref{eq:supvel1} implies that the superfluid velocity vanishes ($\vv_S=0$) and thus Eq.~\eqref{AB4} becomes
\begin{equation} \label{supdens}
    \langle\rhov_n\rangle=(\langle\rho_n\rangle-\rho_S)\vv_N\,.
\end{equation}
The neutron superfluid density can thus be directly inferred from Eq.~\eqref{supdens}, once densities and currents have been computed from the HFB equations.

\section{Results} \label{sec:results}
Before studying the relative flow between superfluid neutrons and clusters in Sec.~\ref{subsec:flow}, let us first discuss the HFB solution in the static case.
\subsection{Band structure and $\beta$-equilibrium}
In the inner crust of neutron stars, nuclear matter is in $\beta$-equilibrium. In terms of the chemical potentials this means
\begin{equation} \label{eq:beta_equilibrium_condition}
    \mu_n=\mu_p+\mu_e\,,
\end{equation}
with $\mu_n$ fixed and $\mu_e$ calculated trough the ultra-relativistic relation
\begin{equation}
    \mu_e = \hbar c (3\pi^2\rho_e)^\frac{1}{3}\,.
\end{equation}
The electron density is determined requiring charge neutrality, i.e. $\rho_e=\langle\rho_p\rangle$.
In contrast to the pasta phases \cite{Almirante24,Almirante24A}, in the 3D crystal phase the number of protons per cell, $Z$, cannot vary continuously in a quantum-mechanical treatment. This is because protons are deeply bound, which means that they occupy practically flat bands. While in the pasta phases one still has a dependence of the single-particle energies on the continuous momentum component(s) in the non-periodic dimension(s), here the full momentum space is decomposed requiring Bloch boundary conditions (see Appendix \ref{appA}). Thus, the proton number can have only certain fixed values, depending on the structure of the discrete energy bands and their degeneracy (if one neglects pairing, as we do for the protons). This is what usually happens in nuclear structure calculations, where the single-particle energies determine shell closures. 

If the neutron chemical potential $\mu_n$ is fixed, one consequence of the above discussed $Z$ discretization is that one can vary the volume $V$ of the direct lattice primitive cell only in a small range without losing $\beta$-equilibrium.

This is due to the fact that the discrete structure of proton bands introduces a ``degeneracy'' in $\mu_p$: if the highest occupied and the lowest unoccupied bands are separated by an energy difference $\Delta\xi=\xi_\beta-\xi_\alpha$, any $\mu_p$ such that $\xi_\alpha<\mu_p<\xi_\beta$ will give the same number of protons $Z$. Hence, when changing the volume $V$ and thus $\rho_e = Z/V$ and the corresponding electron chemical potential $\mu_e$, one stays in $\beta$-equilibrium only as long as $\mu_n-\mu_e$ lies between $\xi_\alpha$ and $\xi_\beta$ (which depend also on $V$).

This means that the procedure usually applied in the pasta phases or in semi-classical treatments (i.e. using the cell volume $V$ that minimizes the thermodynamic potential \cite{Martin15,Yoshimura24}), should here be performed over the different possible $Z$ and the corresponding $\beta$-equilibrium cell volumes. However, here we are not interested in the determination of the composition of the crust, and thus we will explore different possible values of $Z$ at their corresponding $\beta$-equilibrium volumes $V$, in order to check how strongly this choice affects our results for the superfluid fraction.
Concerning the cell volume $V$, we find that the continuous variability interval is within $\simeq 1\fm$ in the distance between clusters, thus we will not explore it in this work.

\begin{figure}
    \centering
    \includegraphics{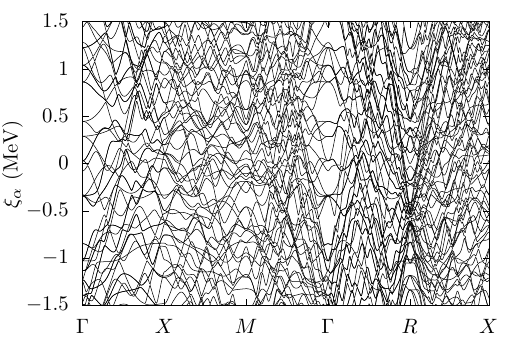}
    \includegraphics{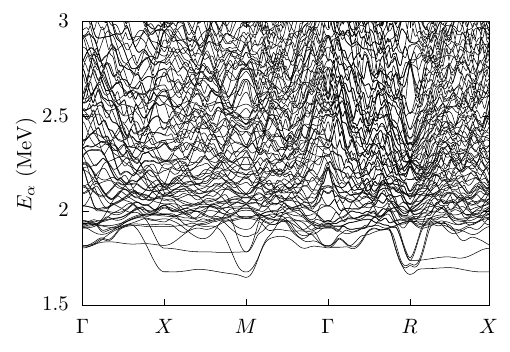}
    \caption{Single-particle (\textit{top}) and quasi-particle (\textit{bottom}) band structure for SLy4 in the simple cubic lattice for $\rho_b=0.033\fm^3$ and $L=26\fm$. $\xi_\alpha=\epsilon_\alpha-\mu$ are HF eigenvalues, first $70$ bands around the Fermi energy are plotted. $E_\alpha$ are HFB eigenvalues, first $90$ positive bands are plotted. The horizontal axis is the Bloch momentum along the high symmetrical path of the simple cubic Brillouin zone \cite{Setyawan10}. In the quasi-particle band structure one can clearly see the pairing gap, in this case its average value is about $2\MeV$.}
    \label{fig:bands}
\end{figure}

Regarding neutron bands instead, since there are a lot more neutrons than protons, many more bands will be occupied. Moreover, neutrons also form a gas around the cluster, which means that many of these bands will have highly non-trivial dependence on the Bloch momentum. As a consequence, in the single-particle spectrum many bands can reach the Fermi energy in different points of the Brillouin zone. In Fig.~\ref{fig:bands}, a portion of both the single-particle (i.e., eigenvalues of $h-\mu_n$ without pairing) and quasi-particle (i.e., eigenvalues of the HFB matrix $\calH$ including pairing) band structure are shown for an example case.

As it can be seen, for the neutrons one has to compute a huge number of bands, and their corresponding eigenvectors. Here we have shown just some tens of them, but in the actual calculations one has to include few thousands: in the HF case, at least as many as the neutron number, in the HFB case even more, because the pairing gap smears the occupation numbers.

In the quasi-particle band structure one can clearly see the pairing gap as the completely empty energy range between the first shown band and its corresponding negative value. Notice that, in the present case, the average value of the gap resulting from the self-consistent calculation is about $2\MeV$, but the pairing interaction we are using here does not result in a uniformly distributed gap over all the bands, and thus some bands feel it less while others more.

\subsection{Density profiles, mean fields, and pairing gaps}

\begin{figure}
    \centering
    \includegraphics[width=8.5cm]{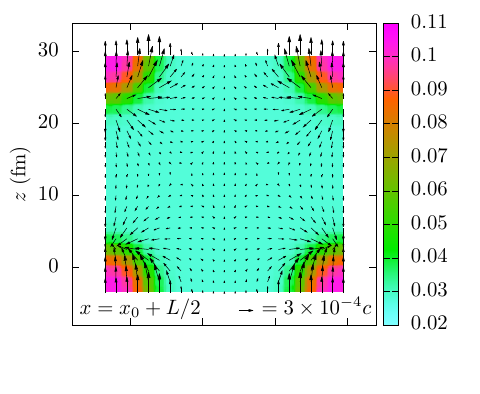}
    \\ \vspace{-16mm}
    \includegraphics[width=8.5cm]{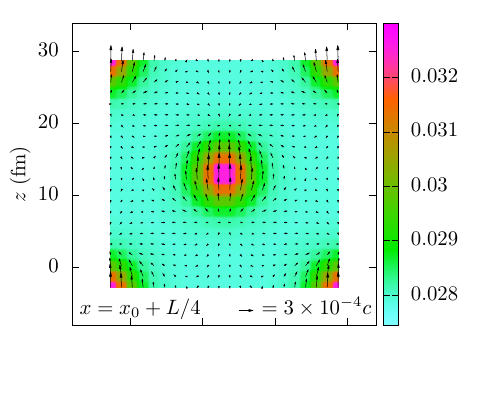}
    \\ \vspace{-16mm}
    \includegraphics[width=8.5cm]{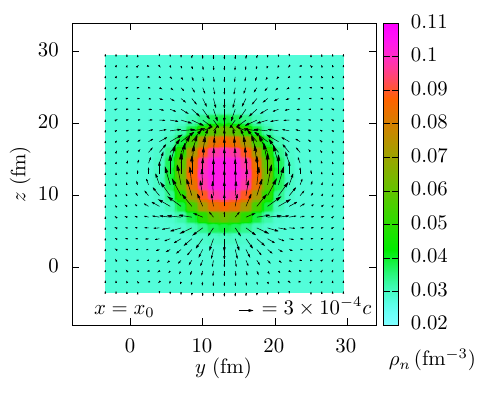}
    \caption{Neutron density (\textit{colors}) and coplanar components of the velocity field (\textit{arrows}) in three sections through the body-centered cubic cell at $x=x_0$ (position of the central cluster, \textit{bottom}), $x=x_0+L/4$ (plane between the clusters, \textit{center}) and $x=x_0+L/2$ (cell boundary, \textit{top}), computed for SLy4, proton number $Z=20$, chemical potential $\mu_n=9\MeV$, baryon density $\rho_b=0.033\fm^{-3}$, bcc cell extension $L=33 \fm$, velocity $\vv_N/c=10^{-3}\zvhat$.}
    \label{fig:densityvelocity3}
\end{figure}

\begin{figure}
    \centering
    \includegraphics{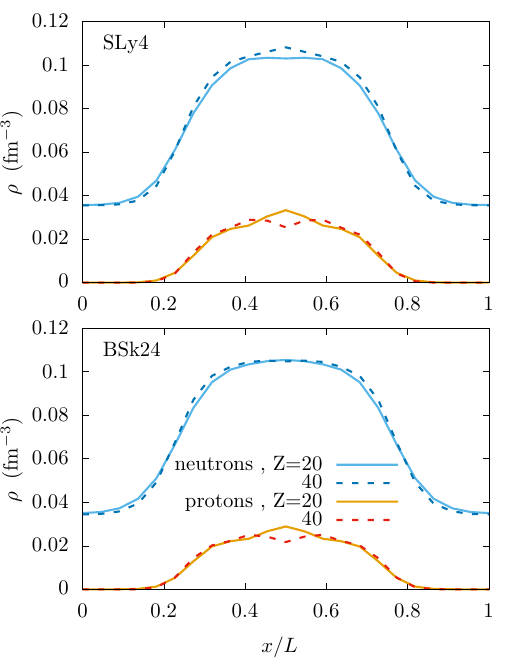}
    \caption{Simple cubic lattice section at $y=z=L/2$ of neutron and proton densities. Results for interactions SLy4 (\textit{top}) and BSk24 (\textit{bottom}) are shown for the same $\rho_b=0.043\fm^{-3}$, cell extension $L=24\fm$ for $Z=20$ and $L=30\fm$ for $Z=40$. The corresponding neutron chemical potentials are $\mu_n^{\text{SLy}}=10\MeV$ and $\mu_n^{\text{BSk}}=9.3\MeV$. For an easier comparison of the cases $Z=20$ and $Z=40$, the $x$ axis is rescaled relative to the cell extension.}
    \label{fig:densities3}
\end{figure}

\begin{figure}
    \centering
    \includegraphics{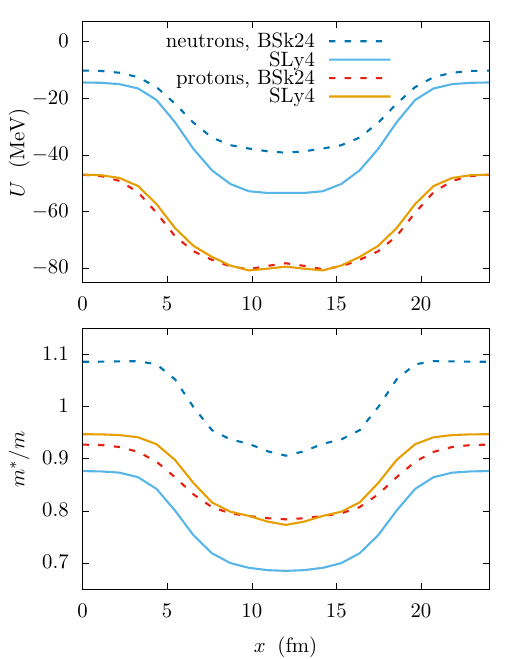}
    \caption{Simple cubic lattice section at $y=z=L/2$ of neutron and proton mean-field potentials $U$ (\textit{top}) and microscopic effective masses $m^*$ (\textit{bottom}), for the same parameters as in Fig.~\ref{fig:densities3} for the case $L=24 \fm$, $Z=20$.}
    \label{fig:meanfield3}
\end{figure}

In Fig.~\ref{fig:densityvelocity3}, neutron density profiles (and coplanar velocity fields, to be discussed in the next Subsection), obtained with the SLy4 interaction are shown for the body-centered cubic (bcc) lattice. We also performed calculations in simple cubic (sc) geometry, but when using the same interaction, proton number, chemical potential and cell volume, we found practically identical density profiles and other relevant quantities such as effective mass and pairing field as in the bcc geometry. As it was found in the rod phase for square and hexagonal lattices, this is due to the fact that physical parameters are the same, and that the clusters are well separated from each other and have a spherical shape. In order to have the same cell volume $V$ in the sc and bcc lattices, we choose the lattice spacings such that
\begin{equation}
    V =  L_{\text{sc}}^3=\frac{L_{\text{bcc}}^3}{2}\,.
\end{equation}
When switching from the SLy4 to the BSk24 interaction, we can readjust the chemical potential to largely compensate for the differences in the mean fields, thereby maintaining the same baryon density. In this way we obtain again comparable density profiles, with a slightly increased size of the cluster in the BSk24 case. In Fig.~\ref{fig:densities3} we show some results for the densities, while in Fig.~\ref{fig:meanfield3} there are the mean-field quantities.

To study the dependence on different numbers of protons $Z$, we consider the allowed values $Z=20$ and $Z=40$, corresponding to magic numbers in the absence of spin-orbit coupling \cite{RingSchuck}. The $\beta$-equilibrium condition is maintained by changing the cell volume such that the average proton density at the different $Z$ is the same.

For both the explored interactions, the neutron and proton density profiles are comparable. Notice that in Fig.~\ref{fig:densities3} the $x$-axis is rescaled with the size of the cubic cell $L$, the volume of the cell for different $Z$ being $V_{Z=40}\simeq2\times V_{Z=20}$. This means that when density profiles at different $Z$ look similar, the cluster with $Z=40$ occupies twice the volume of the one with $Z=20$. This scaling is not surprising, since for a given baryon density in $\beta$-equilibrium, the densities in the gas and in the cluster as well as the volume fraction occupied by the cluster are essentially fixed by infinite-matter properties, as one can see, e.g., in a simple phase-coexistence picture \cite{Martin15}.

The proton density profiles highlight the shell structure of the proton bands. Since protons are deeply bound, the first proton single-particle bands have the same structure as the harmonic oscillator, but with the degeneracy lifted by the fact that, as one can see in the top panel of Fig.~\ref{fig:meanfield3}, the potential is not actually harmonic \cite{RingSchuck}. As a consequence, in the $Z=20$ case, the last occupied level is the $2s$, resulting thus in a seemingly gaussian distribution around the center of the cluster. In the $Z=40$ case instead, the last occupied level is the $2p$, and the proton density has a small dip at the center.  

However, in this work we are not taking into account the spin-orbit interaction, which is of crucial importance to get the correct nuclear compositions (i.e. nuclear energy levels). Thus our results for the shell closures cannot be considered definitive.

In the range of density explored here ($\rho_b\simeq0.03-0.05 \fm^{-3}$), the main difference between the two functionals in the mean-field quantities is in the microscopic effective mass $m^*$. Our results in the neutron gas are in agreement with the effective masses obtained with the same functionals in uniform pure neutron matter \cite{Duan24}, where at these densities $m^*/m$ is expected to be around one for BSk24 and less than one for SLy4, cf. bottom panel of Fig.~\ref{fig:meanfield3}.

\begin{figure}
    \centering
    \includegraphics{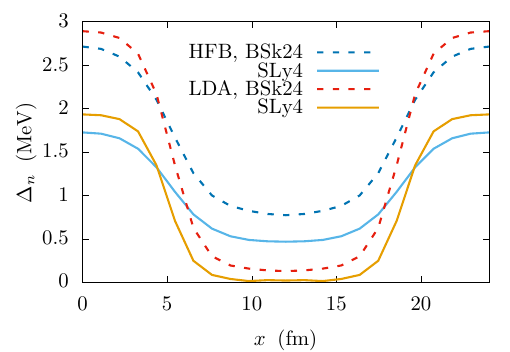}
    \caption{Simple cubic lattice section at $y=z=L/2$ of the neutron pairing gap at the local Fermi momentum as a function of the pair c.o.m. position, for the same parameters as in Fig.~\ref{fig:meanfield3}.}
    \label{fig:gaps}
\end{figure}

Let us now discuss the pairing gaps, shown in Fig.~\ref{fig:gaps}.
As it was the case in the rod phase \cite{Almirante24A}, the gaps are larger in the gas (at the cell boundaries) than in the cluster (in the middle of the cell).
One finds this already at the Local-Density Approximation (LDA) level, i.e. the gap computed at the local density, effective mass and chemical potential.
Our results for the HFB and LDA gaps are presented in Fig.~\ref{fig:gaps} as the blue and red/orange lines, respectively, both evaluated at the local Fermi momentum, i.e. $Q=(3\pi^2\rho_n(\xv))^{1/3}$.
The fact that the BSk24 gaps (dashed lines) are larger than the SLy4 ones (solid lines) reflects the different effective masses.

For both interactions, we see that the difference between the gap in the gas and in the cluster predicted by the LDA is partially washed out in HFB. 
Similar behaviors were found in \cite{Chamel10} when comparing the LDA gaps with the simpler BCS approximation.
For both interactions, the discrepancy between HFB and LDA gaps in the gas is of the order of 10\%, while the gap in the cluster is much bigger in HFB than predicted by LDA due to the so-called proximity effect.

One way to qualitatively understand these results is to look at the ratio between the coherence length $\xi$ of the pair \cite{DeBlasio97} and the typical length scales on which the density varies in space, i.e., the cell extension $L$ and the cluster radius $R$. If $\xi$ is much smaller than the typical lengths of density variations, the correlations between paired particles are less affected by the presence of inhomogeneities. The condition for the validity of the LDA can thus be written as
\begin{equation}\label{eq:coherencelength}
    \xi \ll R, L
    \quad\text{with}\quad
    \xi = \frac{\hbar^2 k_F}{\pi\Delta m^*}\,.
\end{equation}
We can evaluate the above condition for the cases shown in Figs.~\ref{fig:densities3}, \ref{fig:meanfield3}, and \ref{fig:gaps} for $L=24$ and $Z=20$, in which case the cluster radius is approximately $R\simeq 6-7$~fm. In the gas, one has for the coherence lengths $\xi_{\text{SLy}}\simeq8$~fm and $\xi_{\text{BSk}}\simeq5$~fm, which are comparable to $R$ but smaller than $L$. In the cluster instead, one gets $\xi_{\text{SLy}}\simeq50$~fm and $\xi_{\text{BSk}}\simeq25$~fm, which are both greater than $R$ and $L$.
In particular, for the SLy4 interaction, the LDA gap in the cluster is only about $0.05\MeV$, while the HFB one has a value of about $0.5\MeV$, a difference of one order of magnitude.
This implies that in general, in order to evaluate the neutron pairing gap in the neutron-star crust, taking into account the inhomogeneity beyond LDA is of crucial importance.

It has to be noticed that the difference in the pairing gap only slightly affects the densities for both the explored interactions. This is because the gap remains at least one order of magnitude smaller than the mean field potential.

\subsection{Superfluid flow and phase of the gap}\label{subsec:flow}
We now turn to the inclusion of a relative flow between the nuclear clusters and the surrounding superfluid neutrons. Fig.~\ref{fig:densityvelocity3} shows, in addition to the density profile, the coplanar components of the neutron velocity field $\vv_n(\xv)=\rhov_n(\xv)/\rho_n(\xv)$ in the frame where the cluster is moving and the superfluid, in average, carries no momentum. Apart from effects due to momentum-dependent terms in the Skyrme functional, this corresponds to the rest frame of the superfluid.

As seen in Fig.~\ref{fig:densityvelocity3}, neutrons inside the cluster move in the direction of $\vv_N$, i.e., along with the protons, while a counterflow develops in the region outside the cluster. This behavior mirrors what was already observed in the case of the pasta phases \cite{Almirante24,Almirante24A}. Because the system is in three dimensions, the velocity field also includes components perpendicular to $\vv_N$, which can point either toward or away from the cluster depending on the position.

\begin{figure}
    \centering
    \includegraphics[width=8.5cm]{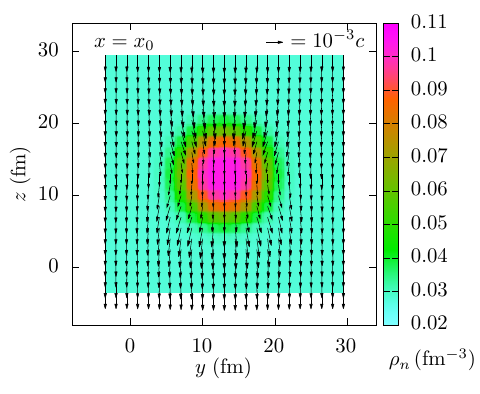}
    \caption{Neutron density (\textit{colors}) and coplanar components of the velocity field (\textit{arrows}) in the plane $x=x_0$ (position of the central cluster) for the same parameters as in Fig.~\ref{fig:densityvelocity3} but in the rest frame of the cluster.}
    \label{fig:densityvelocity3_boost}
\end{figure}

This intricate flow pattern becomes easier to interpret in the rest frame of the cluster, which can be obtained by applying a boost of $-\vv_N$. 
We show this boosted frame picture in Fig.~\ref{fig:densityvelocity3_boost}. 
In that frame, the neutron velocity is roughly $-\vv_N$ $(= -10^{-3}c\,\zvhat)$, the neutron gas flows slightly faster than the neutrons inside the cluster, and the flow tends to pass through the cluster rather than around it. Notice that the velocity scale here is larger by about a factor of 3 than the one in Fig.~\ref{fig:densityvelocity3}.

\begin{figure}
    \centering
    \includegraphics{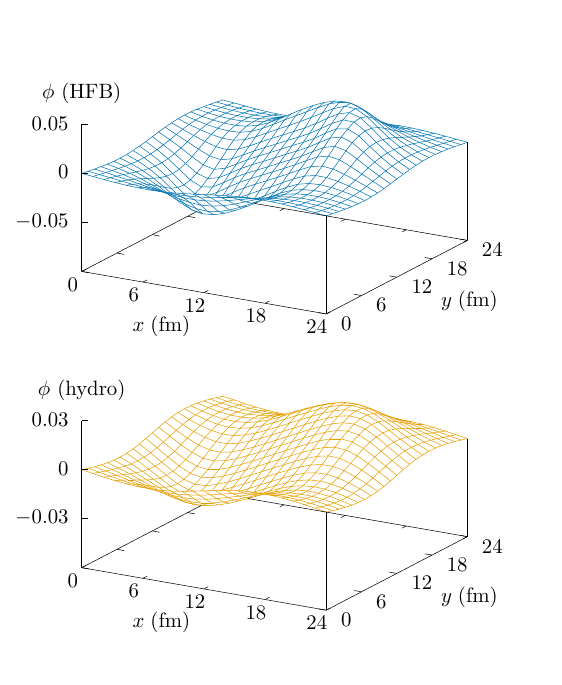}
    \caption{Simple cubic lattice section at $z=L/2$ of the phase of the pairing field, computed for BSk24, proton number $Z=20$, chemical potential $\mu_n=9.3\MeV$, baryon density $\rho_b=0.043 \fm^{-3}$, cell extension $L=24 \fm$, velocity $\vv_N/c=10^{-3}(\xvhat+\yvhat)$. The top panel shows the phase obtained from the HFB calculation, while the bottom panel shows the corresponding result from superfluid hydrodynamics.}
    \label{fig:phase3}
\end{figure}

In Fig.~\ref{fig:phase3}, we show the behavior of the phase of the pairing field, this time for the simple cubic lattice. Our HFB results (upper panel) are in qualitative agreement with the ones obtained in \cite{Martin16} in the framework of superfluid hydrodynamics, with the difference that in our case the cluster has no sharp surface unlike the schematic model used in \cite{Martin16}. To allow for a more direct comparison, we extended the hydrodynamic approach to
incorporate the smooth neutron densities obtained from the HFB calculations, using the same method as in \cite{Saslow76,Saslow12,Rabec25}. This method essentially consists in solving the continuity equation for a given density profile under the assumption that the neutron velocity is entirely determined by the phase of the gap, see Appendix~\ref{appD}. With this extension, the phase obtained from superfluid hydrodynamics (lower panel), has a similar shape as the one obtained from the HFB calculation, but it is smaller in magnitude. When artificially changing the pairing coupling constant (cf. next subsections), we find that the HFB phase approaches the hydrodynamic result only in the limit of large pairing gaps.

\subsection{Superfluid fraction}
As explained in Sec.~\ref{subsec:two-fluid-flow}, averaging the computed neutron current $\rhov_n$ over the cell and using Eq.~\eqref{supdens}, we can determine the superfluid density $\rho_S$ and hence the superfluid fraction $\rho_S/\langle\rho_n\rangle$. One has to perform the calculation for sufficiently small cluster velocity when the system is in the linear response regime, i.e., when $\rho_S$ does not depend on $\vv_N$. To ensure this, we tested different values for the velocity and checked that the current and the phase were proportional to the imposed cluster velocity $|\vv_N|$. We also tested different directions of the flow to check the isotropy of the response. Our results are collected in Table~\ref{3betaequilibrium}.  

The presented results are obtained for $Z=20$ in the simple cubic lattice. We are not showing the bcc ones since at equal baryon density $\rho_b$ and cell volume $V$ they are practically identical to the simple cubic case.
In the BSk24 case, the superfluid fraction is slightly higher than the SLy4 one at comparable baryon density because of the larger pairing gap, as it was the case for the rod phase \cite{Almirante24A}.
Results for a different proton number $Z=40$ are collected in Table~\ref{3betaequilibrium2}. As it can be seen, the superfluid fraction is very slightly increased (within 1\%). Since the pairing gap has almost the same value, and the difference in the neutron densities outside and inside the cluster are comparable (at least in the BSk24 case), this small difference in the superfluid fraction could be ascribed to the decreased ratio between the coherence length $\xi$ and the radius of the cluster in the $Z=40$ case, which brings the system closer to the hydrodynamical limit.

It has to be noticed that the pairing interaction we are using here does not take into account screening effects. These can in general make the value of the neutron pairing gap smaller \cite{Ramanan21}. Thus, in order to check how strongly our results depend on the value of the gap, we report in Table~\ref{varying_gap_SLy4} the superfluid fraction for a chosen configuration, but changing manually the pairing coupling constant $g$, and thus obtaining different values for the pairing gap.
As it can be seen, when the gap is one fourth of our physical value, the superfluid fraction gets reduced by about $10\%$ only. This can be understood in the following way: if one thinks of the superfluid hydrodynamics result as the limiting case for strong gaps, the superfluid fraction will reach a plateau after a certain value of the gap. This is in agreement with what we found in the rod phase, performing a similar analysis \cite{Almirante24A}, and with results in the BCS approximation, which we will discuss in the next Subsection.

\begin{table}
  \caption{\label{3betaequilibrium} 
    Results for baryon density $\rho_b$, average neutron gap at the local Fermi momentum $\langle\Delta_n\rangle$, superfluid fraction $\rho_S/\langle\rho_n\rangle$, and proton fractions $Y_p$ for different chemical potentials $\mu_n$, simple cubic cell extensions $L$ and interactions (SLy4 and BSk24) for $Z=20$.}
  \begin{center}
    \begin{tabular}{ccccccc}
    	\hline
                         int.&$\mu_n$&$L$ &$\rho_b$&$\langle\Delta_n\rangle$&$\rho_S/\langle\rho_n\rangle$&$Y_p$\\
                             &(MeV)  &(fm)&(fm$^{-3}$)&(MeV)&(\%)&(\%)\\\hline
        &{9.0}&26&0.0334&2.03&92.07&3.41\\
        {SLy4}&{10.0}&24&0.0425&1.62&92.84&3.40\\
        &{11.0}&23&0.0518&1.16&94.12&3.37\\ \hline
        &{8.64}&26&0.0342&2.86&92.69&3.33\\
        {BSk24}&{9.30}&24&0.0427&2.55&93.57&3.39\\
        &{9.96}&23&0.0515&2.21&95.25&3.30\\\hline
    \end{tabular}
  \end{center}
\end{table}
\begin{table}
  \caption{\label{3betaequilibrium2} 
    Same as in Table~\ref{3betaequilibrium} but for $Z=40$.}
  \begin{center}
    \begin{tabular}{ccccccc}
    	\hline
                         int.&$\mu_n$&$L$ &$\rho_b$&$\langle\Delta_n\rangle$&$\rho_S/\langle\rho_n\rangle$&$Y_p$\\
                             &(MeV)  &(fm)&(fm$^{-3}$)&(MeV)&(\%)&(\%)\\\hline
        &{9.0}&33&0.0330&2.08&92.70&3.37\\
        {SLy4}&{10.0}&30&0.0423&1.66&93.03&3.40\\
        &{11.0}&29&0.0512&1.22&94.32&3.30\\ \hline
        &{8.64}&33&0.0336&2.87&92.93&3.30\\
        {BSk24}&{9.30}&30&0.0424&2.59&93.76&3.40\\
        &{9.96}&29&0.0510&2.25&95.33&3.32\\\hline
    \end{tabular}
  \end{center}
\end{table}
\begin{table}
  \caption{\label{varying_gap_SLy4} Results for average neutron density $\langle\rho_n\rangle$ and superfluid fraction $\rho_S/\langle\rho_n\rangle$, at different values of the average pairing gap $\langle\Delta_n\rangle$, obtained varying the pairing coupling constant in the interval $g/g_\text{phys}\in[0.7,1]$. Computed for SLy4 at $Z=20$, baryon density $\rho_b=0.033\fm^{-3}$ and simple cubic cell extension $L=26\fm$.}
  \begin{center}
    \begin{tabular}{cccccc}
    	\hline
        $g/g_\text{phys}$&$\langle\Delta_n\rangle$&$\langle\rho_n\rangle$&$\rho_S/\langle\rho_n\rangle$&$Y_p$\\
        & (MeV)  &(fm$^{-3}$)&(\%)&(\%)\\\hline
        0.7&0.54&0.0326&81.17&3.36\\
        0.8&0.93&0.0325&87.08&3.37\\
        0.9&1.44&0.0324&90.17&3.39\\
        1.0&2.03&0.0322&92.07&3.41\\\hline
    \end{tabular}
  \end{center}
\end{table}

\subsection{Comparison with linear response and superfluid hydrodynamics}
The expression for the superfluid density derived in normal band theory \cite{Chamel05,Chamel06,Chamel12} resulted in a very strong entrainment, i.e., small superfluid fraction, when evaluated at densities comparable to the ones explored in this work. Attempts to include pairing did not change this result \cite{Carter05A,Chamel25A}. However, in our recent work \cite{Almirante25}, we found that including the previously neglected ``geometric'' contribution to the linear response expression, one gets a strong increase of the superfluid density with increasing value of the pairing gap.

Here we want to check the validity of the linear response formula, comparing it with the full HFB calculations. To that end, we implement the same physical conditions as in \cite{Almirante25}. We take the mean field potential from the work of Oyamatsu \cite{Oyamatsu93,Oyamatsu94}, where neutron and proton densities are given together with an energy-density functional. Mean field potentials are obtained by taking the derivatives of the latter and performing a Gaussian smearing (to account for the finite range of the nuclear interaction). There is no effective mass and the spin-orbit term is neglected. Moreover, the mean field is treated as a fixed external potential.

Hence, we perform HFB calculations for neutrons keeping the mean field potential from \cite{Oyamatsu94} fixed, while doing the self-consistency in the pairing channel. Different values of the gap can be explored by reducing artificially the pairing coupling constant $g$ (as done in Table~\ref{varying_gap_SLy4}). In this way we can directly compare the HFB superfluid fraction $\rho_S/\langle\rho_n\rangle$ with the one obtained in \cite{Almirante25} within linear response on top of BCS with a constant gap. In fact, here we also evaluate our linear response formula for gaps greater than the ones explored in \cite{Almirante25}, in order to compare results up to the gap obtained with the physical pairing coupling constant $g_\text{phys}$ in the full HFB calculation.

\begin{figure}
    \centering
    \includegraphics{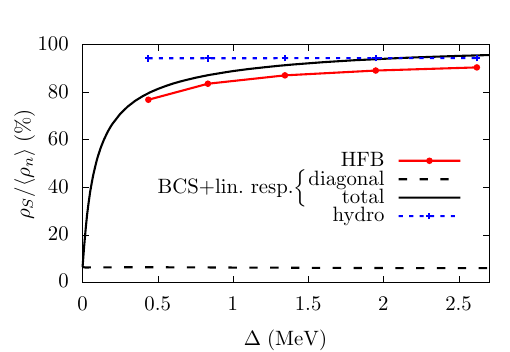}
    \caption{Superfluid fraction $\rho_S/\langle\rho_n\rangle$ for baryon density $\rho_b=0.03\fm^{-3}$ and cell extension $L=34.75 \fm$ (using the mean field potentials of \cite{Oyamatsu93,Oyamatsu94}) computed for different values of the pairing gap $\Delta$. In the HFB case (red points), $\Delta=\langle\Delta_n\rangle$ is the average pairing gap evaluated on the Fermi surface, resulting from the self-consistent calculations when varying the pairing coupling constant in the interval $g/g_{\text{phys}}=[0.6,1]$. At $g=g_{\text{phys}}$ we get $\langle\Delta_n\rangle=2.62\MeV$. In the BCS+linear response case, $\Delta$ is the constant pairing gap, the black solid line is the linear response result Eq.~(15) of \cite{Almirante25}, and the dashed line is instead just the contribution of the diagonal elements ($\alpha=\beta$) to the sum in Eq.~(15) of \cite{Almirante25} (= formula given in \cite{Carter05A,Chamel25A}). The 
blue points correspond to the superfluid fraction obtained in the hydrodynamic description using the HFB density profiles (see Appendix~\ref{appD}).}
    \label{fig:img_supdens_OY}
\end{figure}

Our results are collected in Fig.~\ref{fig:img_supdens_OY}. In the full HFB case, we do not explore smaller values of the gap because one would need finer integration spacing to get reliable results.

We also show results from the expression of \cite{Carter05A,Chamel25A} in the absence of the geometric contribution (dashed line), in which case one finds a weak dependence of the superfluid fraction on the value of the pairing gap, and thus a strong entrainment, very close to the one obtained in normal band theory (limit $\Delta\to 0$).

However, when one includes the geometric contribution, the linear response result becomes much closer to the HFB one. Now, it even exceeds the superfluid fraction obtained in HFB. This can be understood as follows: in our linear response formula Eq.~(15) of \cite{Almirante25}, we neglected the effect of the perturbation on the gap, i.e., the phase. As a consequence, the continuity equation is not fulfilled \cite{Almirante24}, and thus the superfluid density can increase above its hydrodynamical value (dotted line). In the HFB calculation instead, the full response is included, and therefore it approaches the hydrodynamic result from below. This observation is consistent with the statement of \cite{Saslow12} that the hydrodynamic result obtained by solving the continuity equation is an upper bound of the superfluid fraction. We finally note that, within the hydrodynamic approach, the superfluid fraction is essentially independent of the pairing gap, since it is entirely determined by the neutron density profile.

Also, in Fig.~\ref{fig:img_supdens_OY} one can see that the HFB results get closer to those obtained in linear response on top of BCS when the magnitude of the pairing is reduced. This could be due to the fact that the smaller is the gap, the better works the BCS approximation with a constant gap.

\section{Conclusions} \label{sec:conclusion}
In this paper we investigated the superfluid properties of the crystal phase in $\beta$-equilibrium at baryon densities $\rho_b\simeq0.03-0.05\fm^{-3}$ as expected in the inner crust of neutron stars. This was achieved by performing HFB calculations with Bloch boundary conditions, using an energy density functional for the mean-field and a non-local separable potential, fitted on $V_{\text{low-}k}$, for the pairing field.

Two different lattice geometries were implemented, namely simple cubic and body-centered cubic. Under the same interaction, baryon density, and cell volume, they produce nearly identical results for neutron and proton density profiles, neutron gap, and superfluid density.

For the mean fields, we used two distinct energy density functionals: SLy4 and BSk24. As expected, the neutron effective masses are higher in BSk24 compared to SLy4, which is consistent with results for pure neutron matter \cite{Duan24}. Consequently, the pairing gap is significantly larger in the BSk24 case.

The most interesting findings relate to the superfluid dynamics. We introduce a relative stationary flow between the clusters and the surrounding superfluid neutron gas, within the linear regime for the relative velocity.
In the superfluid rest frame (in the sense $\vv_S = 0$), we observed that outside the cluster, there is an average counterflow of neutrons opposite to the cluster motion, while inside the cluster, neutrons flow in the same direction as it. This behavior is also evident from the periodicity of the phase of the pairing field.
Additionally, there is some motion in the perpendicular direction.
This is due to the preference of neutrons to flow through the cluster, in the frame in which the cluster is at rest. Our velocity fields are consistent with those from superfluid hydrodynamics. For both the energy density functionals explored, the superfluid fraction at $\rho_b = 0.033\fm^{-3}$ is $\rho_S/\langle\rho_n\rangle\simeq93\%$. This is almost as large as the prediction $\rho_S/\langle\rho_n\rangle\simeq96\%$ from superfluid hydrodynamics, which according to \cite{Saslow12} represents an upper bound for this quantity. 

In conclusion, we find that the entrainment effect in the inner crust is quite weak, with the superfluid fraction exceeding $90\%$ of the average neutron density in layers where previous band structure calculations \cite{Chamel12A,Chamel25A} predicted a superfluid fraction of less than $10\%$. This confirms the finding of our recent analysis within linear-response theory, which under a couple of simplifying assumptions (no effective mass, BCS approximation with a constant gap, neglect of the phase of the gap) allowed us to pinpoint the origin of this discrepancy to the significant role of the geometric contribution to the superfluid density \cite{Almirante25}.

As demonstrated in \cite{Martin16}, this amount of superfluid neutrons in the crust is sufficient to explain the glitches observed in the Vela pulsar, with no need to modify glitch models to account for superfluidity in the core, as suggested in \cite{Andersson12}. Even if the gap was reduced by a factor of two or three due to screening effects, this would not alter this conclusion. This result also has implications for low-lying phonons \cite{Pethick10, Chamel13B, Durel18}, and consequently, for the thermal evolution of neutron stars \cite{Page12}.

\begin{acknowledgments}
We gratefully acknowledge support from the CNRS/IN2P3 Computing Center (Lyon - France) for providing computing resources needed for this work.\end{acknowledgments}

\appendix
\section{HFB with Bloch boundary conditions in 3D} \label{appA}
Here we show how one can rewrite the HFB equations in momentum space such that periodicity and Bloch boundary conditions are taken into account, similarly to what we did in 1D \cite{Almirante24} and 2D \cite{Almirante24A}. One can impose periodicity (details on the particular choices implemented in this work are given in Appendix~\ref{appB}) for the basic quantities, namely normal and anomalous density matrices, i.e.,
\begin{align}
\langle \psi^\dagger_\uparrow(\rv'+\av_i) \psi_\uparrow(\rv+\av_i) \rangle 
  &= \langle \psi^\dagger_\uparrow(\rv')\psi_\uparrow(\rv) \rangle\,,\label{period-rho}\\
\langle \psi_\downarrow(\rv'+\av_i) \psi_\uparrow(\rv+\av_i) \rangle 
  &= \langle \psi_\downarrow(\rv') \psi_\uparrow(\rv) \rangle\,,
  \label{period-kappa}
\end{align}
where $\psi$ is the field operator and $\av_i$ are the primitive vectors of the direct lattice, $i\in\{1,2,3\}$.

For given vectors $\av_i$, the primitive vectors of the reciprocal lattice can be found using the well-known relation \cite{Ashcroft}
\begin{equation} \label{eq:reclat}
    \bv_i\cdot\av_j=2\pi\delta_{ij}\,,
\end{equation}
which will be useful when going to momentum space.

For the anomalous density matrix, one can write
\begin{equation}
\langle c_{-\pv'\downarrow} c_{\pv\uparrow} \rangle
 = \int \! d^3 r\, d^3r' e^{-i\pv\cdot\rv}\, e^{i\pv'\cdot\rv'} \langle \psi_\downarrow(\rv') \psi_\uparrow(\rv) \rangle
   \,,
   \label{eq:ccavg}
\end{equation}
where $c$ is the annihilation operator. Performing a change in the integration variables $(\rv,\rv') \to (\rv+ \av_i ,\rv'+ \av_i)$ and using Eq.~\eqref{period-kappa}, one obtains
\begin{align}
\langle c_{-\pv'\downarrow} c_{\pv\uparrow} \rangle 
  =&\int\! d^3 r\, d^3 r'\, e^{-i\pv\cdot\rv}\, e^{i\pv'\cdot\rv'} \langle \psi_\downarrow(\rv') \psi_\uparrow(\rv) \rangle
  \nonumber \\
  &\times e^{-i 2\pi (p_i-p'_i)}\,,
  \label{eq:ccavgshifted}
\end{align}
where $p_i$ are components of the momentum $\pv$ in the basis $\{\bv_1,\bv_2,\bv_3\}$, i.e., $\pv = \sum_i p_i\bv_i$. Combining Eqs.~\eqref{eq:ccavg} and \eqref{eq:ccavgshifted}, one finds that the momentum labels must satisfy the following conditions: $p_i-p'_i\in\mathbb{Z}$.


Now we rewrite the momentum components in the basis $\{\bv_1,\bv_2,\bv_3\}$ as a sum of an integer $n_i$ and a continuous part $p_{bi}$. In this way, the momentum is decomposed as $\pv = \nv+\pv_b$ where $\nv = \sum_i n_i\bv_i$ is a vector of the reciprocal lattice and $\pv_b = \sum_i p_{bi}\bv_i$ is the Bloch momentum defined in the first Brillouin zone (BZ). Then the condition $p_i-p'_i\in\mathbb{Z}$ implies that the anomalous density matrix is diagonal in the Bloch momentum $\pv_b$ and has non-diagonal elements only in the integer labels $n_i$.
For the normal density matrix one can proceed in a completely analogous way.
As a consequence, also the HFB matrix is diagonal in $\pv_b$.

Summarizing, for each triple $(p_{b1},p_{b2}, p_{b3})$ we have an HFB matrix as in Eq.~\eqref{eq:HFBH} with indices $n_1,n_2,n_3,n'_1,n'_2,n'_3$.
We diagonalize it according to Eq.~\eqref{eq:HFBeigen}, obtaining quasi-particle energies $E_\alpha(\pv_b)$ and eigenvectors $(U^*_{\alpha \nv}(\pv_b),-V^*_{\alpha \nv}(\pv_b))$. In terms of these, the normal and anomalous density matrices are expressed as
\begin{align}
    \rho_{\nv\nv'}(\pv_b)&=\sum_{E_\alpha>0} V_{\nv' \alpha}^*(\pv_b) V_{\nv \alpha}(\pv_b)\,,
    \\
    \kappa_{\nv\nv'}(\pv_b)&=\sum_{E_\alpha>0} U_{\nv \alpha}^*(\pv_b) V_{\nv' \alpha}(\pv_b)\,,
\end{align}    
with $E_\alpha = E_\alpha(\pv_b)$. Then, one can compute the densities and pairing field as
\begin{align} 
    \rho(\xv) =& 2\int_\text{BZ} \frac{d^3p_b}{(2\pi)^3}
    \sum_{\nv \nv'} e^{i(\nv-\nv')\cdot\xv}\,\rho_{\nv \nv'}(\pv_b)\,,
    \label{eq:dens1D}
\\    
    \tau(\xv)=&~2\int_\text{BZ} \frac{d^3p_b}{(2\pi)^3}
    \sum_{\nv \nv'}e^{i(\nv-\nv')\cdot\xv}
    \rho_{\nv \nv'}(\pv_b)
    \nonumber\\
    &\times (\nv+\pv_b)\cdot(\nv'+\pv'_b)\,,
\\
    \jv(\xv) =&~2
    \int_\text{BZ} \frac{d^3p_b}{(2\pi)^3}
    \sum_{\nv \nv'}  e^{i(\nv-\nv')\cdot\xv}
    \rho_{\nv \nv'}(\pv_b)
    \nonumber\\
    &\times\Big(\frac{\nv+\nv'}{2}+\pv_b\Big) \,,
\\ 
    \Delta_{\nv \nv'}(\pv_b)=&~g~f_{\nv+\nv'}(\pv_b) \sum_{\vect{m} \vect{m}'} \delta_{\nv-\nv',\vect{m}-\vect{m}'}
    \nonumber\\
    &\times\int_\text{BZ} \frac{d^3p_b}{(2\pi)^3}
    f_{\vect{m}+\vect{m}'}(\pv_b) \kappa_{\vect{m}\vect{m}'}(\pv_b)\,,
    \label{eq:gapkn}
\\
\intertext{where}
    f_{\nv+\nv'}(\pv_b) =&
    \exp\bigg(-\frac{(\frac{\nv+\nv'}{2}+\pv_b)^2}{k_0^2}\bigg)\,.
\end{align}

All of this is also true for the HF case, with the simplification that there are no anomalous density and pairing field. Thus, instead of the HFB matrix, only the mean-field Hamiltonian $h_{\nv \nv'}(\pv_b)$ needs to be diagonalized. Denoting its eigenvalues and eigenvectors $\epsilon_\alpha(\pv_b)$ and $V_{\alpha \nv}(\pv_b)$, the density matrix reads then
\begin{equation} \label{eq:densmatHF}
    \rho_{\nv \nv'}(\pv_b)=\sum_{\epsilon_\alpha<\mu} V_{\nv' \alpha}^*(\pv_b) V_{\nv \alpha}(\pv_b)\,.
\end{equation}
%
\section{Reciprocal lattice and Brillouin zone} \label{appB}
In Appendix~\ref{appA} we discussed the general formalism for HFB with Bloch boundary conditions in 3D. Since we are working in momentum space, it is important to construct the reciprocal lattice and the BZ such that they respect the symmetries of the direct lattice.

For the simple cubic lattice one has
\begin{align} \label{eq:ksc}
    \begin{cases}
    \av_1=L \xvhat,&
    \bv_1=\frac{2\pi}{L}\xvhat,\\
    \av_2=L\yvhat,&
    \bv_2=\frac{2\pi}{L}\yvhat,\\
    \av_3=L\zvhat,&
    \bv_3=\frac{2\pi}{L}\zvhat,
    \end{cases}
\end{align}
while for the body-centered cubic one
\begin{align} \label{eq:kbcc}
    \begin{cases}
    \av_1=\frac{L}{2}(-\xvhat+\yvhat+\zvhat),&
    \bv_1=\frac{2\pi}{L}(\yvhat+\zvhat),\\
    \av_2=\frac{L}{2}(\xvhat-\yvhat+\zvhat),&
    \bv_2=\frac{2\pi}{L}(\xvhat+\zvhat),\\
    \av_3=\frac{L}{2}(\xvhat+\yvhat-\zvhat),&
    \bv_3=\frac{2\pi}{L}(\xvhat+\yvhat),
    \end{cases}
\end{align}
where $L$ is the extension of the conventional unit cell (which in the bcc case contains two clusters).

The BZ is defined as the Wigner-Seitz cell around a reciprocal lattice point \cite{Ashcroft}. In practice, it can be constructed from the parallelepiped given by $\pv_b = \sum_i p_{bi} \bv_i$ with $p_{bi}\in (-\frac{1}{2},\frac{1}{2}]$ by reshuffling parts of it into equivalent regions with smaller absolute values. To that end, we define for each Bloch momentum $\pv_b$ in the parallelepiped the set of 27 equivalent Bloch momenta given by
\begin{equation}\label{eq:set-equiv-bloch}
    \{\pv_b\} = \left\{\sum\nolimits_i(p_{bi}+\lambda_i) \bv_i\right\} \quad\text{with}\quad \lambda_i\in\{-1,0,1\}.
\end{equation}
Then, the Brillouin zone is obtained by taking the smallest (in absolute value) among these equivalent Bloch momenta.

Since in our numerical calculations we discretize the primitive cell with $N^3$ points, our momentum space will be truncated. Similar to the BZ, also the bounds of our reciprocal lattice must have the symmetry of the reciprocal lattice itself. In order to achieve this, we proceed analogously to the construction of the BZ. We start from the set of reciprocal lattice vectors given by $\nv=\sum_i n_i\bv_i$ 
with $n_i$ integers in the interval $(-N/2,N/2]$.
%
%
For each triple $(n_1,n_2,n_3)$, one can construct the set of 27 reciprocal lattice vectors equivalent to it on the discrete mesh points as 
%
\begin{equation} \label{eq:set-equiv-reciprocal}
    \{ \nv \}=\left\{\sum\nolimits_i (n_i+N\lambda_i)\bv_i\right\} \quad \text{with} \quad \lambda_i\in\{-1,0,1\}.
\end{equation}
%
%
Then the truncated reciprocal lattice used in the calculation is obtained by taking the smallest (in absolute values) of the equivalent reciprocal lattice vectors in $\{\nv\}$.

In the simple cubic lattice, the above procedure is trivial (since the smallest vectors of Eqs. \eqref{eq:set-equiv-bloch} and \eqref{eq:set-equiv-reciprocal} are always those with $\lambda_i = 0$). In the body-centered cubic lattice instead, one starts from a rhombohedron, while after selecting the smallest equivalent vectors as explained above, one gets the correct rhombic dodecahedron shape of the BZ and of the truncated momentum space.

\section{Numerical details}
As shown in Appendix \ref{appA}, our problem is diagonal in the Bloch momentum $\kv_b$, thus we deal with an HFB matrix in the integer momenta $\nv,\nv'$. Since we want to perform the self-consistent calculations on a machine, the problem has to be discretized. We choose to discretize the primitive cell in the three directions $\av_i$ ($i=1,2,3$) with a spacing $\av_i/N$. This naturally introduces a cutoff $n_i \in (-N/2,N/2]$ in the integer coefficients of the reciprocal lattice vectors 
%
%
(before the reshuffling of the reciprocal lattice mentioned in Appendix~\ref{appB}).
As a consequence, the HFB matrix has dimension $2N^3\times2N^3$ (the HF matrix instead only $N^3\times N^3$). In this way, we can access the first $N^3$ bands for both neutrons and protons. In this work, we take $N=22$.

For each Bloch momentum $\kv_b$, a diagonalization of both the HFB and the HF matrices in momentum space is performed. The relevant quantities to construct the matrices are computed in coordinate space and then Fourier transformed.

For $\kv_b$ of neutrons, we take $N_b^{\text{sc}}=216$ points in the first Brillouin zone of the simple cubic lattice, while $N_b^{\text{bcc}}=234$ for the body-centered cubic lattice.

For protons, since they are strongly confined, only a few bands are occupied and they are flat. As an approximation one could replace the integration over the Brillouin zone with the product between the BZ volume and the integrand computed in $\kv_b=0$.

Unfortunately performing the calculation in this way gives rise to spurious results in the currents. This is a discretization effect coming from the asymmetry in the integer momenta bounds. Taking as example the simple cubic lattice, one has in the three directions $n_i\in(-N/2,N/2]$. If the $N/2$ mode has some contribution to the current (although it should be small), computing it in $\kv_b=0$ would give a term that, even when there should be no current, will not be cancelled by the corresponding term in $-N/2$. In order to avoid this effect, we choose for both protons and neutrons integration points in the Brillouin zone such that none of their components is zero and, for each integration point, there are also those obtained from it by implementing the symmetry transformations of the reciprocal lattice (i.e. in the simple cubic lattice by flipping one by one the signs of its components). Then, if one of the integer momentum components $n_i=N/2$, the BZ is split into two parts: for $k_{bi}<0$, we keep $n_i=N/2$, while for $k_{bi}>0$, we replace it by $n_i=-N/2$. In this way parity is respected at the BZ level and there is no spurious effect.

For the reasons explained above, in the case of protons, we take in the BZ $N_b^{\text{sc}}=8$ points for the simple cubic lattice, and $N_b^{\text{bcc}}=12$ points for the body-centered cubic lattice.

Finally, it has to be noticed that it is not necessary to include all the computed energy bands in the calculation of the densities. For both neutrons and protons we have at our disposal $N^3$ bands. In the case of protons, since they are normal fluid only, one includes as many bands as the number of protons (accounting for the spin degeneracy), and actually this is the way in which one can fix the proton number $Z$. In the case of neutrons instead, the presence of pairing smears the occupation numbers and thus more bands have to be included. In practice we include the first 6000 bands, which for the analyzed cases correspond to an energy cutoff of the order of $100\MeV$.

For all the choices we discussed about discretization, BZ integration points and energy cutoff, we have verified that if we increase them the self-consistent calculations converge to the same results. Our convergence criterion is defined such that for each point in the cell
\begin{align}
    |\rho^{(m)} - \rho^{(m+1)}| &< |\rho^{(m+1)}|\times10^{-4},\nonumber\\
    |j^{(m)} - j^{(m+1)}| &< |j^{(m+1)}|\times10^{-4},\nonumber\\
    |\Delta_0^{(m)} - \Delta_0^{(m+1)}| &< |\Delta_0^{(m+1)}|\times10^{-4},
\end{align}
where $\rho^{(m)}$ is the result of the $m$-th iteration etc.

In order to speed up the convergence, we use Broyden's modified method as discussed in \cite{Baran08} and already used in the HFB description of the pasta phases in \cite{Almirante24,Yoshimura24,Almirante24A}. Our Broyden vector has dimension $11N^3 + 2$ and it is defined as ($U_n(\xv)$, $U_p(\xv)$, $\hbar^2/2m^*_n(\xv)$, $\hbar^2/2m^*_p(\xv)$, $\Iv_n(\xv)$, $\Iv_p(\xv)$, $\Delta^0_{\nv-\nv'}$, $\mu_n$, $\mu_p$). $\Delta^0_{\nv-\nv'}$ is the gap defined in Eq.~\eqref{eq:gapkn} without the factor $g~f_{\nv+\nv'}(\kv_b$). We performed the method using the results of $M=3$ previous iterations and a mixing coefficient $\alpha=0.7$.
\section{Superfluid hydrodynamics with arbitrary density profile}
\label{appD}

In this appendix, we present the calculation of the superfluid fraction within the framework of hydrodynamics, following the method introduced
by Saslow~\cite{Saslow76}. We work in the conventional unit cell of the lattice.

The neutron superfluid is described by the pairing gap given in
Eq.~(\ref{eq:DeltaQx}), and the neutron velocity field is determined by
the phase $\phi(\rv)$ according to~\cite{Martin16}
\begin{equation}
    \vv_n(\rv) = \frac{\hbar}{2m_n}\,\nablav \phi(\rv),
    \label{eq:vhydro}
\end{equation}
which is the central assumption of the superfluid hydrodynamics approach. We note that the factor 2 which is absent in \cite{Saslow76} appears because the phase of the gap involves the phases of two single-particle wave functions.

Similarly to Sec.~\ref{subsec:two-fluid-flow}, we work in the frame where the superfluid is at rest and the clusters move with velocity ${\vv}_N$. In this frame, the phase of the pairing field can be taken as a periodic function over the conventional unit cell~\cite{Martin16},
\begin{equation}
    \phi(\rv + L \hat{\mathbf{e}}_i) = \phi(\rv),
    \qquad (i = x,y,z),
    \label{periodicity}
\end{equation}
where we have assumed a cubic cell of size $L$, and $\hat{\mathbf{e}}_i$ denotes the unit vector along the $i$-th direction. The choice of the conventional unit cell instead of the primitive one is only made for simplicity, the resulting phase reflects the periodicity of the density profile and hence it is also periodic with respect to the primitive lattice.

The continuity equation for neutrons reads
\begin{equation}
    \frac{\partial \rho_n(\rv, t)}{\partial t} 
    + \nablav\cdot \rhov_n(\rv, t) = 0,
    \label{continuity}
\end{equation}
where the time-dependent neutron density is defined as:
\begin{equation}
    \rho_n(\rv, t) = \rho_n(\rv - \vv_N t),
    \label{density_shift}
\end{equation}
obtained by shifting the static density profile along the direction of the cluster velocity $\vv_N$. The corresponding neutron current is:
\begin{equation}
    \rhov_n(\rv, t) = \rho_n(\rv, t)\, \vv_n(\rv, t).
    \label{current}
\end{equation}

Substituting Eqs.~\eqref{density_shift} and \eqref{current} into the continuity equation~\eqref{continuity} and evaluating at $t=0$ gives
\begin{equation}
    \rho_n(\rv)\, \nablav\!\cdot\!\vv_n(\rv)
    + \nablav \rho_n(\rv) \cdot \vv_n(\rv)
    = \vv_N \cdot \nablav \rho_n(\rv)
    \label{ode1}
\end{equation}
or, using Eq.~\eqref{eq:vhydro},
\begin{equation}
    \rho_n(\rv)\,\nablav^2 \varphi(\rv)
    + \nablav \rho_n(\rv)\cdot \nablav \varphi(\rv)
    = \vv_N \cdot \nablav \rho_n(\rv).
    \label{ode2}
\end{equation}
where the scalar velocity potential $\varphi(\rv)$ is related to the phase $\phi(\rv)$ by
\begin{equation}
    \varphi(\rv) = \frac{\hbar}{2m_n}\, \phi(\rv).
    \label{velocity_field}
\end{equation}

We solve the second-order equation~\eqref{ode2} for $\varphi(\rv)$ using the Fourier spectral method, which takes advantage of the periodicity of
the field. The scalar potential $\varphi(\rv)$ and the
neutron density $\rho_n(\rv)$ are expanded as
\begin{equation}
    \varphi(\rv) = \sum_{\qv} \varphi_{\qv}\, e^{i\qv\cdot\rv},\qquad
    \rho_n(\rv) = \sum_{\qv} \rho_{n,\qv}\, e^{i\qv\cdot\rv},
\end{equation}
where the discrete wavevectors in a cubic box of length $L$ are given by $\qv = \frac{2\pi}{L}(n_x,n_y,n_z)$, with $n_x,n_y,n_z \in \mathbb{Z}$. In Fourier space products become convolutions, leading to the linear system
\begin{equation}
    \sum_{\qv^{\,\prime}} A_{\qv\qv^{\,\prime}}\, \varphi_{\qv^{\,\prime}}
    = i (\qv\cdot\vv_N)\, \rho_{n,\qv},
    \label{fourier eq}
\end{equation}
with matrix elements
\begin{equation}
    A_{\qv\qv^{\,\prime}}
    = -\rho_{n,\qv - \qv^{\,\prime}}\, (\qv \cdot \qv^{\,\prime}).
    \label{A}
\end{equation}

Using the neutron densities obtained from the HFB calculation,
we solve the linear system for the coefficients $\varphi_{\qv}$. The
real-space field $\varphi(\rv)$ then follows from the inverse transform,
and the neutron velocity is given by
$\vv_n(\rv) = \nablav \varphi(\rv)$. The superfluid fraction is finally
obtained from Eq.~\eqref{supdens}.

\bibliography{refs}
\end{document}